\documentstyle[preprint,prl,aps,psfig,floats,epsf,multicol]{revtex}
\newcommand{\be}{\begin{equation}}
\newcommand{\ee}{\end{equation}}
\newcommand{\bea}{\begin{eqnarray}}
\newcommand{\eea}{\end{eqnarray}}

\renewcommand{\L}{{\Lambda_{QCD}}}
\newcounter{dafigcounter}
\newcommand{\pfig}[3]{
 \refstepcounter{dafigcounter}
 \begin{minipage}[t]{#2}
  \begin{center}
   {\epsfxsize=#2 \mbox{\epsffile{#1.eps}}}
  \end{center}
  \label{#1}
  \small \bf Fig.~\thedafigcounter\rm\ #3
 \end{minipage}
}
\begin{document}
\draft
%\twocolumn
%
\tighten
\firstfigfalse
%\twocolumn[\hsize\textwidth\columnwidth\hsize\csname@twocolumnfalse\endcsname
%
\title{Leading-Order Gluon-Pair Production from a Space-Time 
Dependent Chromofield}
\author{Dennis D. Dietrich, Gouranga C. Nayak, and Walter Greiner}
\address
{\small\it{Institut f\"ur Theoretische Physik,
J. W. Goethe-Universit\"at,
60054 Frankfurt am Main, Germany}}
%

%\date{\today} 
%{\tiny\verb$Id: sig.tex,v 2.4 1999/03/08 23:29:03 misha Exp $}}
\maketitle    

%%%%%%%%%%%%%%%%%%%%%%%%%%%%%%%%%%%%%%%%%%%%%%%%%%%%%%%%%%%%%%%%%%%%%%%%%%%%%%%
\begin{abstract}

We describe gluon-pair production from a space-time dependent chromofield via 
vacuum polarization within the framework of the background field method of 
QCD. The processes we consider are first order in the action. We derive the 
corresponding source terms for gluon pair production for situations that can 
be described by a 1+1 dimensional approach. Especially, we 
observe that within the range of applicability of our approach the 
principal contribution to gluon production is included. Gluon production from 
a space-time dependent chromofield will play an important role in the 
production and evolution of the quark-gluon plasma in ultra relativistic 
heavy-ion collisions at RHIC and LHC.

\end{abstract}

%\bigskip

%
%\narrowtext

%%%%%%%%%%%%%%%%%%%%%%%%%%%%%%%%%%%%%%%%%%%%%%%%%%%%%%%%%%%%%%%%%%%%%%%%%%%%%%%

\twocolumn

\section{Introduction}

The quark-gluon plasma (QGP), a deconfined state of matter, is predicted to 
exist at high temperatures around $200MeV$ and/or high energy densities 
around $2GeV/fm^3$ \cite{lattice}. Apart from situations shortly after the 
big bang or inside a neutron star, heavy-ion colliders could produce the 
necessary conditions for its existence \cite{qm98}. Even if this state of 
matter is then available in the laboratory, it will live for only such minute 
timespans and inside so small volumes that its direct detection is rendered 
impossible. This is the reason, why one is forced to look for indirect 
signatures like $J/\Psi$ suppression \cite{satzs}, strangeness enhancement 
\cite{rafelski,singh}, and electromagnetic probes such as dileptons and photons
\cite{strickland,shuryak,alam,nayak-dilep}. But, due to many uncertainties in 
the theoretical predictions and difficulties in the experiments, it is up 
to now still not possible to conclude 
about the existence of the QGP based on observations of these indicators. 
That fact is mostly due to the deficient knowledge about the space-time 
evolution of the parton distribution in these reactions. Initially, they are 
in pronounced non-equilibrium. The way, the QGP relaxes towards equilibrium 
has got to be studied carefully. This can in principle be achieved through 
the solution of a relativistic non-abelian transport equation for quarks and 
gluons. Herefore, it is necessary to understand the production of partons in 
ultra-relativistic heavy-ion collisions. There, the two colliding nuclei 
travel almost at the speed of light and are thus highly Lorentz contracted. 
When they pass through each other, a chromofield is formed between them due to 
the exchange of soft gluons \cite{all1,all2,all3,more,nayak,gcn}. This 
picture is 
the extension of the color flux-tube or string model widely used in high 
energy $pp$, $e^+e^-$, and $pA$ collisions \cite{coll,lund}. The chromofield 
so formed polarizes the QCD vacuum and produces $q\bar{q}$-pairs and gluons. 
These partons collide with each other and proceede towards equilibrium.

In all previous studies \cite{all2,nayak,gcn}, the production of the QGP is 
investigated by taking parton production from a constant chromofield into 
account (which has been studied in \cite{schwinger,casher}). However, as seen 
in numerical studies \cite{nayak,gcn}, the chromofield acquires a strong 
space-time dependence due to a combination of such effects as expansion, 
background acceleration, color rotation, parton collision and parton 
production. In situations like these, parton production from a constant 
chromofield is not applicable and one has got to find the corresponding 
expression for a general space-time dependent field.

$e^+e¯$ production from a space-time dependent chromofield is studied by 
Schwinger \cite{schwinger}. This is equivalent to computing the probability 
from the amplitude $<k_1,k_2|S^{(1)}|0>$ (denoted by 
$A_{cl}\rightarrow e^+(k_1)e^-(k_2)$) with $S^{(1)}$ obtained from the 
interaction lagrangian density of a classical field and quantized Dirac fields 
\cite{izju}. Due to the same structure of the interaction lagrangian density, 
the production of a $q\bar{q}$ pair is similar to the $e^+e^-$ case except for 
color factors \cite{nayak2}. However, the computation of the probability for 
the production of gluons from a space-time dependent chromofield is greater 
than that for the production of fermionic, massive quarks and antiquarks. 
That is due to the larger phase space at the disposal for particles described 
by  the adjoint representation. Hence, for a correct study of the production 
and equilibration of the QGP, it is necessary to describe, how gluons are 
produced from a strongly variable chromofield.

In \cite{ddd} we have studied for the first time, the production of gluon 
pairs from a space-time dependent chromofield via vacuum polarization. There, 
we derived corresponding source terms, discussed their range of 
applicability, and explicitely evaluated them for a time-dependent model field.
However, the chromofield formed at RHIC and LHC is not only time dependent, 
but inhomogeneous in space and time. In the pre-equilibrium stage of the 
evolution 
of the QGP it is a good approximation to consider a 1+1-dimensionally 
expanding system, {\it i.e} $A$ depends on the time $t$ and the longitudinal 
coordinate $z$. In this paper, we derive the source term for a $t$ and $z$ 
dependend chromofield. Also, for a comparison to Schwinger's 
formula for 
fermion-pair production (see Eq. 6.33 in \cite{schwinger}), we evaluate 
the total probability for gluon-pair production from a space-time dependent 
chromofield. In future, we hope to use the source terms derived in this paper
for a $t$ and $z$ dependend chromofield in a non-abelian relativistic 
transport equation to study the production and evolution of the QGP for 
situations similar to those found at RHIC and LHC.

Finally, we discuss the validity of our approach to parton production,
especially in view of an application to situations found at RHIC and LHC.
As our approach is perturbative, it is only capable of describing the 
production of particles with a momentum $p$ larger than the product of the 
coupling constant $g$ and the gauge field $A$. This is the principal 
contribution to the production of particles as long as $gA$ is larger than 
the hadronic scale of QCD, {\it i.e.} $\L$. So, we can use our method when the 
condition $p>gA>\L$ is satisfied. That means also that our approach breaks 
down as soon as the decaying field $A$ reaches the value $\L/g$ from above. 
After this time $t_{end}$, a non-perturbative treatment of particle 
production is needed which is beyond the scope of this paper.
In this paper, we investigate, whether the stage of 
particle production is already almost over before we reach $t_{end}$ or 
whether any 
significant contribution in the semi-hard sector is still to be expected. 

The paper is organized as follows: In chapter II, we explain the setup 
of our method by deriving the probability for the production of gluon pairs 
to the lowest order in the effective action. In section III, we give the 
source terms for a chromofield variable in 1+1 dimensions. Chapter IV 
contains a brief summary.

%%%%%%%%%%%%%%%%%%%%%%%%%%%%%%%%%%%%%%%%%%%%%%%%%%%%%%%%%%%%%%%%%%%%%%%%%%%%%%%

\section{Probability for the Production of Gluon Pairs}

In this section, we will compute the probability for the production of gluon 
pairs from a space-time dependent chromofield via vacuum polarization.
The process which
contributes to the gluon pair production in leading order of the action
is diagramatically represented in Fig.(1). To evaluate these diagrams,
we follow the background field method of QCD which, in a gauge invariant 
manner contains a classical
background field and a quantum gluonic field simultaneously. 

In the background field method of QCD \cite{dewitt,thooft}, the gauge field 
$A_{\mu}$ is split into two parts: $A_{\mu}\rightarrow A_{\mu}+Q_{\mu}$.
Here, $A_{\mu}$ is a classical background field which will not be quantized 
and  $Q_{\mu}$ is a quantum field representing the gluons. As the $A$-field 
is not quantized it does not occur in the functional integral and not in the 
coupling to the external source 
$[dA]\rightarrow[dQ]~~~{\rm and}~~~JA\rightarrow JQ$.

In this method, a so called background field gauge: 

\be
G^a=
\partial^{\mu}Q^a_{\mu}+gf^{abc}A^{b\mu}Q^c_{\mu}=
(D^{\mu}[A]Q_{\mu})^a
\label{gauge}
\ee

is chosen in a way as to make the total lagrangian 
${\cal L}={\cal L}_G+{\cal L}_{GF}+{\cal L}_{FP}$
with the lagrangian density for the gauge fields:

\be
{\cal L}_G=-\frac{1}{4}F^a_{\mu\nu}[A+Q]F^{a\mu\nu}[A+Q],
\ee

with the contribution to the total lagrangian due to gauge fixing 
\cite{thooft}

\be
{\cal L}_{GF}=-\frac{1}{2}(G^a)^2,
\ee

and the corresponding lagrangian density for the ghost fields

\be
{\cal L}_{FP}=-(D_{\mu}[A]\chi^{\dagger}_a)(D^{\mu}[A+Q]\chi)_a
\ee

gauge invariant with respect to type I gauge transformations \cite{itz}.
We evaluate the vacuum-polarization matrix element $M=<k_1k_2|S^{(1)}|0>$
where $S^{(1)}$ contains all interaction terms of the Lagrangian density 
involving two $Q$-fields.
For the diagram for the production of two gluons by 
coupling to the $A$-field once see Fig.(1a). 
The Feynman rule for this vertex 
can be obtained from that part of the lagrangian density which involves one 
$A$ field and two $Q$ fields in the corresponding part of the lagrangian 
density and is given by:

\bea
(V_{1A})^{abd}_{\mu\nu\rho}
=
~\nonumber \\
=
gf^{abd}[-2g_{\mu\rho}K_{\nu}
-g_{\nu\rho}(k_1-k_2)_{\mu}
+2g_{\mu\nu}K_{\rho}].
\eea

For the production of two gluons by coupling to the background field twice 
see Fig.(1b). 
The Feynman rule for this vertex 
can be extracted from that part of the lagrangian density which involves two 
$A$ fields and two $Q$ fields in the corresponding part of the lagrangian 
density and is given by:

\bea
(V_{2A})^{abcd}_{\mu\nu\lambda\rho}
=
~\nonumber \\
=
-ig^2
[f^{abx}f^{xcd}
(g_{\mu\lambda}g_{\nu\rho}-g_{\mu\rho}g_{\nu\lambda}+g_{\mu\nu}g_{\lambda\rho})
~\nonumber \\
+f^{adx}f^{xbc}
(g_{\mu\nu}g_{\lambda\rho}-g_{\mu\lambda}g_{\nu\rho}-g_{\mu\rho}g_{\nu\lambda})
~\nonumber \\
+f^{acx}f^{xbd}
(g_{\mu\nu}g_{\lambda\rho}-g_{\mu\rho}g_{\nu\lambda})].
\eea

These Feynman rules coincide with those derived in \cite{abbott}. The 
amplitude for the production of gluon-pairs from the processes shown in 
Fig.(1) is given by:

\be
M=M_{1A}+M_{2A}
\ee

where

\bea
M_{1A}=
\frac{(2\pi)^2}{2}\int d^4K\delta^{(4)}(K-k_1-k_2)
~\nonumber \\
A^{a\mu}(K)\epsilon^{b\nu}(k_1)\epsilon^{d\rho}(k_2)(V_{1A})^{abd}_{\mu\nu\rho}
\label{ampy}
\eea

for the three-vertex and:

\bea
M_{2A}=
\frac{1}{4}\int d^4k_3 d^4k_4\delta^{(4)}(k_1+k_2-k_3-k_4)
~\nonumber \\
A^{a\mu}(k_3)A^{c\lambda}(k_4)\epsilon^{b\nu}(k_1)\epsilon^{d\rho}(k_2)
(V_{2A})^{abcd}_{\mu\nu\lambda\rho}
\label{ampx}
\eea

for the four-vertex. Here $A(K)$ is the 
Fourier transformation of the classical field $A(x)$ given by:

\be
A(K)=\frac{1}{(2\pi)^2}\int d^4x e^{+iK\cdot x}A(x).
\ee

Note that the above amplitudes include all the weight factors needed in order 
to retrieve the correct lagrangian density. The probability is obtained 
from the amplitudes by the relation:

\be
W=\sum_{spin}
\int\frac{d^3k_1}{(2\pi)^32k_1^0}\frac{d^3k_2}{(2\pi)^32k_2^0}MM^*.
\label{prob}
\ee

To obtain the correct physical gluon polarizations in the final state we use:

\be
\sum_{spin}\epsilon^{\nu}(k_1)\epsilon^{*\nu'}(k_1)=
\sum_{spin}\epsilon^{\nu}(k_2)\epsilon^{*\nu'}(k_2)=-g^{\nu\nu'}
\ee

for the spin-sum of the gluons and afterwards deduct the corresponding ghost 
contributions 
which are shown in Fig.(2). For the integration over the phase space of the 
gluons there is an additional factor of $1/2$ because these particles (not 
the ghosts) in the final state are identical. Evaluating Eq.(\ref{prob}) for 
the gluons, not yet including the ghost corrections, we find:

\be
W^A=W_{1A,1A}+W_{1A,2A}+W_{2A,1A}+W_{2A,2A}
\label{wa}
\ee

with:

\bea
W_{1A,1A}&=&\frac{10}{8}\alpha_S\int d^4K
\times
~\nonumber \\
&\times&
~[(A^a(K)\cdot A^{*a}(K))K^2
-
~\nonumber \\
&&-
(A^a(K)\cdot K)(A^{*a}(K)\cdot K)],
\label{w1a}
\eea

\bea
W_{1A,2A}=W_{2A,1A}=\frac{3ig\alpha_S}{4}\int d^4K\frac{d^4k_3}{(2\pi)^2}
~\nonumber \\
f^{aa'c'}((A^a(K)\cdot A^{*a'}(k_3))(A^{*c'}(K-k_3)\cdot K)),
\label{w12a}
\eea

and:

\bea
W_{2A,2A}=\frac{\alpha_Sg^2}{16}
\int\frac{d^4k_3}{(2\pi)^2}\frac{d^4k'_3}{(2\pi)^2}d^4K
T^{aca'c'}_{\mu\lambda\mu'\lambda'}
~\nonumber \\
~[((A^{a\mu}(k_3)A^{c\lambda}(K-k_3))
\times
~\nonumber \\
\times
   (A^{*a'\mu'}(k'_3)A^{*c'\lambda'}(K-k'_3)),
\eea

with:

\bea
T^{aca'c'}_{\mu\lambda\mu'\lambda'}
&=&
g_{\mu\lambda}g_{\mu'\lambda'}
\times
~\nonumber \\
&&\times
(f^{ab,cd}+f^{ad,cb})(f^{a'b,c'd}+f^{a'd,c'b})
+
~\nonumber \\
&+&
12f^{ac,a'c'}g_{\mu\mu'}g_{\lambda\lambda'},
\eea

with:

\be
f^{ab,cd}=f^{abx}f^{xcd}.
\ee

It has to be noted that the area of integration for $K$ is limited to 
$(K)^2>0$ and $K^0>0$, because real gluon pairs are to be produced. Further, 
despite of a factor of $i$, Eq.(\ref{w12a}) is real.

Now, we calculate the ghost part.
The vertices involving two ghosts and one 
classical field and two ghosts and two classical fields respectively can be 
read from the lagrangian density and are given by:

\be
(V^{FP}_{1A})^{abd}_{\mu}=gf^{abd}(k_1-k_2)_{\mu}
\ee

and:

\be
(V^{FP}_{2A})^{abcd}_{\mu\lambda}=
-ig^2g_{\mu\lambda}(f^{ab,cd}+f^{ad,cb}).
\ee

The corresponding amplitudes for the ghosts are given by:

\be
(M^{FP})^{bd}=(M^{FP}_{1A})^{bd}+(M^{FP}_{2A})^{bd}
\ee

with:

\bea
(M_{1A}^{FP})^{bd}=\frac{(2\pi)^2}{2}\int d^4K
\times
~\nonumber \\
\times
\delta^{(4)}(k_1+k_2-K)A^{a\mu}(K)(V^{FP}_{1A})^{abd}_{\mu}
\eea

and:

\bea
(M_{2A}^{FP})^{bd}
&=&
\frac{1}{4}\int d^4k_3 d^4k_4
\times
~\nonumber \\
&&\times
\delta^{(4)}(k_1+k_2-k_3-k_4)
\times
~\nonumber \\
&&\times
A^{a\mu}(k_3)A^{c\lambda}(k_4)(V^{FP}_{2A})^{abcd}_{\mu\lambda}.
\eea

The probability in this case is: 

\be
W^{FP}=
\int\frac{d^3k_1}{(2\pi)^32k_1^0}\frac{d^3k_2}{(2\pi)^32k_2^0}
(M^{FP})^{bd}(M^{FP})^{*bd}.
\label{probfp}
\ee

Evaluating Eq.(\ref{probfp}) yields:

\be
W^{FP}=W^{FP}_{1A,1A}+W^{FP}_{2A,2A}
\label{wfp}
\ee

with:

\bea
&&W^{FP}_{1A,1A}=
-\frac{\alpha_S}{8}\int d^4K
\times
~\nonumber \\
&\times&
~[(A^a(K)\cdot A^{*a}(K))K^2
-
~\nonumber \\
&&-
(A^a(K)\cdot K)(A^{*a}(K)\cdot K)]
\label{gh1}
\eea

and:

\bea
&&W^{FP}_{2A,2A}=
\frac{\alpha_Sg^2}{32(2\pi)^4}\int d^4K
\times
~\nonumber \\
&\times&
(A^a(K)*A^c(K))(A^{a'}(K)*A^{c'}(K))
\times
~\nonumber \\
&&\times
(f^{ab,cd}+f^{ad,cb})(f^{a'b,c'd}+f^{a'd,c'b}).
\label{gh2}
\eea

Note that $W_{1A,2A}^{FP}=W_{2A,1A}^{FP}=0$.
To obtain the probability $W$ for the production of a real $gg$ pair from a
space-time dependent classical chromofield $A$, the result for the ghosts 
from Eq.(\ref{wfp}) has to be subtracted from the result for the gluons given in
Eq.(\ref{wa}).

%%%%%%%%%%%%%%%%%%%%%%%%%%%%%%%%%%%%%%%%%%%%%%%%%%%%%%%%%%%%%%%%%%%%%%%%%%%%%%%

\section{Chromofield in 1+1 dimensions}

We now give the source term for gluon production for a $t$ and $z$ dependent  
chromofield. This 1+1 dimensional approach is likely to propely describe
the very early phase of a heavy ion-collision at RHIC and LHC.
The probability for particle production and the source term are connected by:

\be
W=\int d^4xd^3k\frac{dW}{d^4xd^3k}.
\ee

Note that of any expression given for a source term throughout the paper only 
the real part has got to be taken into account. 
The source term for the production of gluon pairs already corrected with the 
respective ghost contribution again can be composed analogously to
Eq.(\ref{wa}).
The general expressions for the gluon source terms given in \cite{ddd} in 
1+1 dimensions specialize to:

\bea
&&\frac{dW_{gg}^{1A,1A}}{d^4xd^3k}
=
\frac{3g^2}{16(2\pi)^4k^0}A^{a\mu}(\vec x_L)
\times
~\nonumber \\
&\times&
\int dx'_Le^{i\vec k_L\cdot(\vec x_L-\vec x'_L)} A^{a'\mu'}(\vec x'_L)
\times
~\nonumber \\
&\times&
[
8g_{\mu\mu'}-3(k_{\mu}k_{\mu'}+k'_{\mu}k'_{\mu'})
-
~\nonumber \\
&&-
5(k_{\mu}k'_{\mu'}+k'_{\mu}k_{\mu'})
]_{\vec k'_L\rightarrow-i\vec\partial_L}^{\vec k'_T\rightarrow-\vec k_T}
\times
~\nonumber \\
&\times&
K_0(|\vec k_T|\zeta)
\label{dw11}
\eea 

where the transverse momentum components of $k'$ are replaced by derivatives 
with respect to $ix_T$. In principle, they can be calculated analytically, but 
the result is rather lengthy. Vectors with index $L$ or $T$ are always defined 
as follows:

\be
\vec x_L=(t,0,0,z)
{\rm~and~}
\vec k_T=(0,k_x,k_y,0)
\ee

Additionally, we define the {\sl distance} $\zeta$ as:

\be
\zeta=(-(x_L-x'_L)^2)^{\frac{1}{2}}.
\ee

The remaining contributions are:

\bea
&&\frac{dW_{gg}^{2A1A}}{d^4xd^3k}
=
\frac{3g^3}{8(2\pi)^4k^0}A^{a'\mu'}(\vec x^L)
\times
~\nonumber \\
&\times&
\int \frac{d^2\vec x'_L}{\zeta}A^{a\mu}(\vec x^{'L})A^{c\lambda}(\vec x^{'L})
f^{a'ac}g_{\mu\mu'}
\times
~\nonumber \\
&\times&
e^{i\vec k_L\cdot(\vec x'_L-\vec x_L)}K_1(|\vec k_T|\zeta)
(\vec k_L-i\vec x_L)_{\lambda}
\label{dw21}
\eea

and

\bea
&&\frac{dW_{gg}^{2A2A}}{d^4xd^3k}
=
\frac{g^4}{64(2\pi)^2}A^{a\mu}(\vec x_L)A^{c\lambda}(\vec x_L)
\times
~\nonumber \\
&\times&
\int d^2\vec x'_L 
A^{a'\mu'}(\vec x'_L)A^{c'\lambda'}(\vec x'_L)
\times
~\nonumber \\
&\times&
e^{i[\vec k_L\cdot(\vec x_L-\vec x'_L)]}K_0(|\vec k_T|\zeta)
\tilde T^{aca'c'}_{\mu\lambda\mu'\lambda'},
\label{dw22}
\eea

where $\tilde T$ differs from $T$ by an additional factor of 2 in the second
summand. The total source term is the sum of Eq.(\ref{dw11}), twice
Eq.(\ref{dw21}), and Eq.(\ref{dw22}). 
Now, we ask the question, if the processes depicted in Figs. \ref{Fig2} and 
\ref{Fig3}
mainly only contribute to the particle multiplicities inside the range of 
applicability of our approach. For this purpose we choose a special form of 
the field:

\be
A^{a\mu}(x)=A^{a3}_{in}e^{-t/t_0}\theta(t+z)\theta(t-z)\theta(t),
\ee

which is confined to the inside of the light-cone. All other components are 
equal to zero. Many other forms of the field could have been chosen, but this
choice is inspired by a numerical study presented in \cite{gcn}. We define the 
time $t_f$ which $gA$ needs to reach $\L$:

\be
t_f=t_0\ln\frac{gA_{in}}{\L}
\ee

For $t_0 = 0.5 fm$, $g=1.5$, $A_{in}=1.5GeV$, and $\L=150MeV$ $t_f$ is 
approximately equal to $1.3fm$. The values for the decay time $t_0$, the 
coupling constant $g$, and the initial magnitude of the gauge field $A_{in}$ 
are rough estimates for an ultra-relativistic heavy-ion collision at LHC (see 
\cite{ddd}). Now let us regard the accumulated density of produced particles 
up to a time $t$:

\be
w=\int_0^t dt' \frac{dW}{d^3\vec xdt'}.
\ee

In our case this quantity is given by:

\bea
w
=
\frac{g^2A^2_{in}}{16\pi t_0}
[
11(e^{-2\frac{|z|}{t_0}}-e^{-2\frac{t}{t_0}})
+
~\nonumber \\
+
36g^2A^2_{in}t^2_0(e^{-4\frac{|z|}{t_0}}-e^{-4\frac{t}{t_0}})
]
\times
~\nonumber \\
\times
\theta(t+z)\theta(t-z)\theta(t).
\label{w_}
\eea

In Fig.(\ref{dwt}) we have plotted the accumulated density $w$ up to the 
time $t_f$ for the parameters given above and for different values of the 
longitudinal coordinate $z$. One observes right 
away that the principal contribution to the production of particles is from 
times significantly smaller than $t_f$. To quantify this, let us concentrate 
on the accumulated number-density $w$ at $t_f$ in the central region $z=0$. We 
find: 

\bea
w
=
\frac{1}{16\pi t_0}
[
11(g^2A^2_{in}-\L^2)
+
~\nonumber \\
+
36t^2_0(g^4A^4_{in}-\L^4)
].
\eea

From Eq.(\ref{w_}) one sees that for $gA_{in}>>\L$ the typical time to 
approach this end value of the density is of the order $t_0$. Which is by a
factor of $\ln(gA_{in}/\L)$ smaller than the time $t_f$. So for the set of 
parameters chosen here, this factor is approximately $2.7$. Again regarding
Fig.(\ref{dwt}), we see that this argumentation is well justified. From these 
facts also arises that the produced particle density is sharply peaked in the 
central region (see Fig.(\ref{dwz})). This is because the size of the expanding
system is still comparably small when most of the particles are produced.

%%%%%%%%%%%%%%%%%%%%%%%%%%%%%%%%%%%%%%%%%%%%%%%%%%%%%%%%%%%%%%%%%%%%%%%%%%%%%%%

\section{Summary}

To summarize, we have derived the probability for the production of a gluon 
pair from an arbitrary space-time dependent chromofield to the first oder in 
the effective action within the framework of the background field method of 
QCD. In order to make a connection to the experimental situation, we have 
derived a source term for the production of gluons from a $t$ and $z$ 
dependent external chromofield. For this 1+1 dimensional description of the 
plasma in the pre-equilibrium stage of an ultra-relativistic heavy-ion 
collision, we have observed that the principal contribution to the production 
of gluons is included within the range of applicability of our approach.
Also, for a comparison with Schwinger's result for the total probability for 
the production of fermion anti-fermion pairs, we have evaluated the total 
probability for the production of real gluon pairs from an arbitrary
space-time dependent chromofield. 
We plan to include the source terms derived in this paper in a non-abelian 
transport equation with collisions between the partons taken into account. 
A selfconsistent solution of this equation would serve 
to study the production and equilibration of the quark-gluon plasma at RHIC 
and LHC.

%%%%%%%%%%%%%%%%%%%%%%%%%%%%%%%%%%%%%%%%%%%%%%%%%%%%%%%%%%%%%%%%%%%%%%%%%%%%%%%

\section*{Acknowledgements}

The authors want to thank
Stefan Hofman, Chung-Wen Kao, Alexander Krasnitz, Larry McLerran, 
Dirk Rischke, 
Raju Venugopalan, and Qun Wang for their helpful discussions.
D.D.D. and G.C.N. want to thank the Graduiertenf\"orderung des Landes 
Hessen and the Alexander von Humboldt Foundation respectively for financial 
support.

%%%%%%%%%%%%%%%%%%%%%%%%%%%%%%%%%%%%%%%%%%%%%%%%%%%%%%%%%%%%%%%%%%%%%%%%%%%%%%%

\onecolumn

%%%%%%%%%%%%%%%%%%%%%%%%%%%%%%%%%%%%%%%%%%%%%%%%%%%%%%%%%%%%%%%%%%%%%%%%%%%%%%%
\begin{figure}[thb]
\begin{center}
\pfig{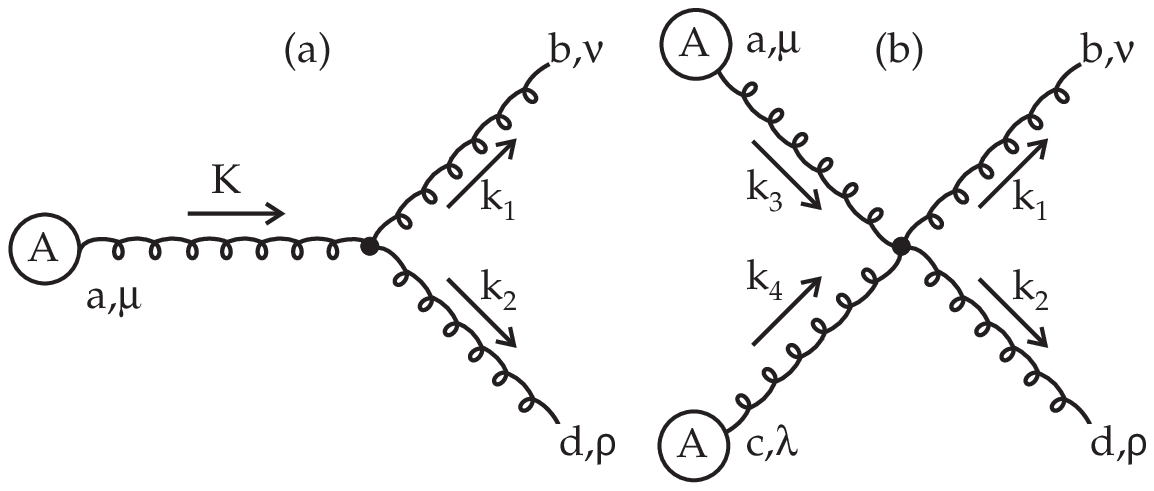}{12cm}
{
Feynman diagrams for the production of two gluons by coupling to the 
field $A$ once or twice
}
\end{center}
\end{figure}
%%%%%%%%%%%%%%%%%%%%%%%%%%%%%%%%%%%%%%%%%%%%%%%%%%%%%%%%%%%%%%%%%%%%%%%%%%%%%%%
\begin{figure}[thb]
\begin{center}
\pfig{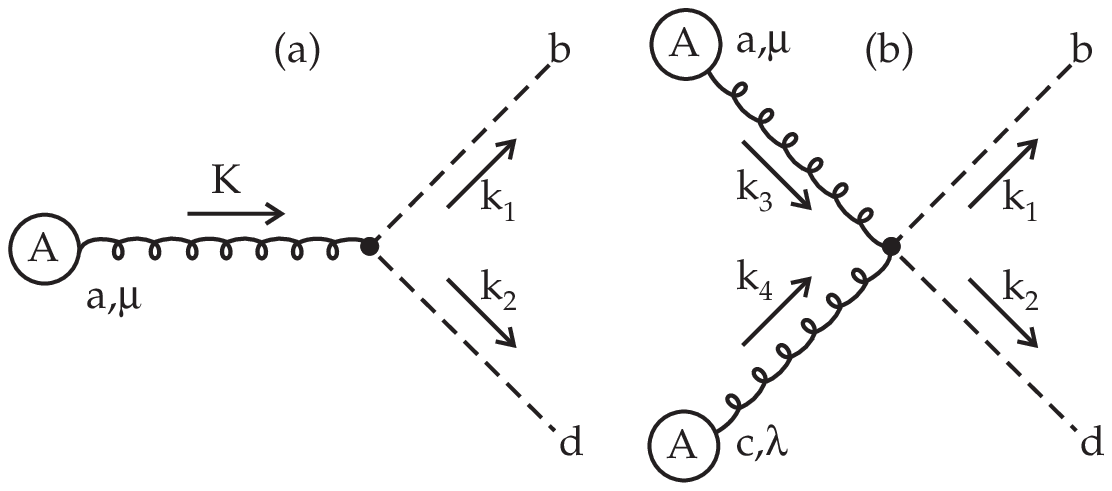}{12cm}
{
Feynman diagrams for the ghosts, corresponding to the gluon vertices.
}
\end{center}
\end{figure}
%%%%%%%%%%%%%%%%%%%%%%%%%%%%%%%%%%%%%%%%%%%%%%%%%%%%%%%%%%%%%%%%%%%%%%%%%%%%%%%
\begin{figure}[thb]
\begin{center}
\pfig{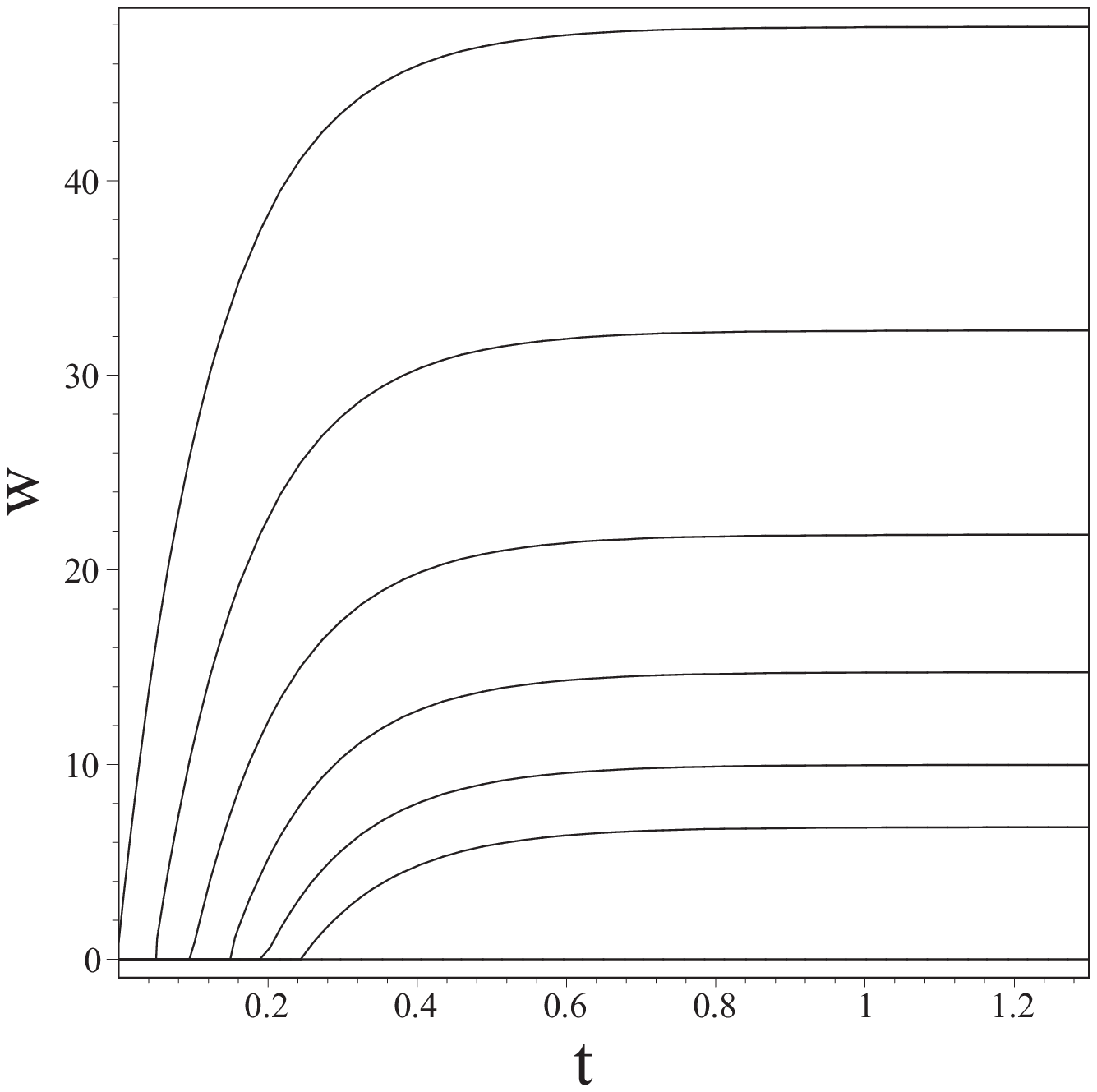}{12cm}
{Accumulated density of particles $w$ in $GeV^3$ versus the elapsed time $t$ 
in $fm$ for different 
values of the longitudinal coordinate $z$. From top to bottom these are: 
$z=0.00fm$, $z=0.05fm$, $z=0.10fm$, $z=0.15fm$, $z=0.20fm$, and $z=0.25fm$.}
\end{center}
\end{figure}
%%%%%%%%%%%%%%%%%%%%%%%%%%%%%%%%%%%%%%%%%%%%%%%%%%%%%%%%%%%%%%%%%%%%%%%%%%%%%%%
\begin{figure}[thb]
\begin{center}
\pfig{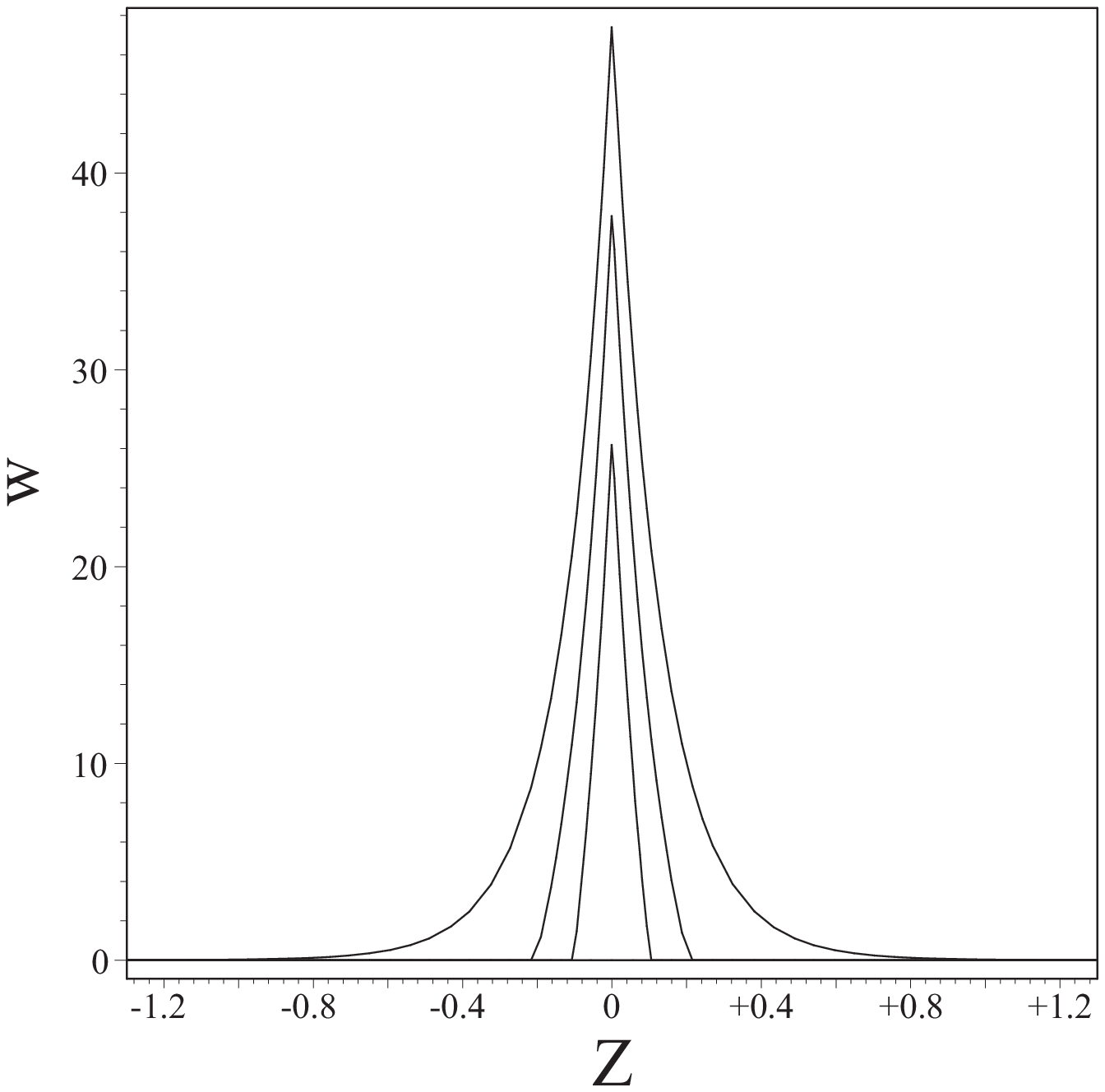}{12cm}
{Accumulated density of particles $w$ in $GeV^3$ versus the longitudinal 
coordinate $z$ in $fm$ for different values of the elapsed time $t$ in $fm$. 
From top to bottom these are: $t=1.3fm$, $t=0.2fm$,and $t=0.1fm$.}
\end{center}
\end{figure}
%%%%%%%%%%%%%%%%%%%%%%%%%%%%%%%%%%%%%%%%%%%%%%%%%%%%%%%%%%%%%%%%%%%%%%%%%%%%%%%

\end{document}